\documentclass[aps, prd, preprint, superscriptaddress, tightenlines, nofootinbib, showpacs]{revtex4}

\usepackage{amssymb, amsmath} 
\usepackage{stmaryrd}

\usepackage{bbm} 

\usepackage[english]{babel}
\usepackage{slashed}
\usepackage{cancel}

\usepackage{slashbox}

\usepackage{graphicx}
\usepackage{color}

\usepackage{datetime}
\settimeformat{ampmtime}

\usepackage{float}

\usepackage[bookmarks,breaklinks,plainpages=false,pdfpagelabels]{hyperref} 
	\hypersetup{linkcolor=red,citecolor=blue,filecolor=dullmagenta,urlcolor=darkblue} 

\usepackage{soul}

\definecolor{grey}{rgb}{0.4,0.4,0.4}
\definecolor{lightgrey}{rgb}{0.6,0.6,0.6}
\definecolor{dullmagenta}{rgb}{0.4,0,0.4}
\definecolor{darkblue}{rgb}{0,0,0.4}
\definecolor{orange}{rgb}{1,0.5,0}
\definecolor{lightbrown}{rgb}{0.75,0.5,0.25}
\definecolor{tan}{cmyk}{0.14,0.42,0.56,0}
\definecolor{djunglegreen}{cmyk}{0.99,0,0.52,0}
\definecolor{lightgreen}{rgb}{0,1,0}
\definecolor{olivegreen}{cmyk}{0.64,0,0.95,0.40}
\definecolor{midgreen}{rgb}{0.0,0.675,0.0}
\definecolor{darkgreen}{rgb}{0,0.5,0}

\newcommand{\q}{\quad}
\newcommand{\qq}{\qquad}

\newcommand{\vs}{\vspace}

\renewcommand{\.}{\hspace{0.5mm}}

\newcommand{\ra}{\ensuremath{\rightarrow}}

\newcommand{\Scal}{\ensuremath{\mathcal{S}}}

\newcommand{\Rbbm}{\ensuremath{\mathbbm{R}}}

\renewcommand{\d}{\ensuremath{\mathrm{d}}}

\newcommand{\eg}{e.g.}
\newcommand{\ie}{i.e.}

\newcommand{\cf}{cf.}

\makeatletter
 \def\ifundefined#1{\expandafter\ifx\csname#1\endcsname\relax}
 \ifundefined{default@color}
  \let\default@color=\current@color
 \fi
\makeatother

\newcommand{\beq}{\begin{equation}}
\newcommand{\eeq}{\end{equation}}
\newcommand{\bea}{\begin{eqnarray}}
\newcommand{\beas}{\begin{eqnarray*}}

\newcommand{\eea}{\end{eqnarray}}
\newcommand{\eeas}{\end{eqnarray*}}

\begin{document}

\title{Decay of Graviton Condensates and their Generalizations\\ in Arbitrary Dimensions}

\author{Florian K{\"u}hnel}
\email{florian.kuhnel@fysik.su.se}
\affiliation{The Oskar Klein Centre for Cosmoparticle Physics,
	Department of Physics,
	Stockholm University,
	AlbaNova,
	106\.91 Stockholm,
	Sweden}

\author{Bo Sundborg}
\email{bo.sundborg@fysik.su.se}
\affiliation{The Oskar Klein Centre for Cosmoparticle Physics,
	Department of Physics,
	Stockholm University,
	AlbaNova,
	106\.91 Stockholm,
	Sweden}

\date{\formatdate{\day}{\month}{\year}, \currenttime}

\begin{abstract}
Classicalons are self-bound classical field configurations, which include black holes in General Relativity. In quantum theory, they are described by condensates of many soft quanta. In this work, their decay properties are studied in arbitrary dimensions. It is found that generically the decays of other classicalons are enhanced compared to pure graviton condensates, \ie~black holes. The evaporation of higher dimensional graviton condensates turns out to match Hawking radiation solely due to non-linearites captured by the classicalon picture. Although less stable than black holes, all self-bound condensates are shown to be stable in the limit of large mass. Like for black holes, the effective coupling always scales as the inverse of the number of constituents, indicating that these systems are at critical points of quantum phase transitions. Consequences for cosmology, astro- and collider physics are briefly discussed.
\end{abstract}

\pacs{04.40.-b, 04.50.Gh, 14.70.Kv, 04.50.-hq}

\maketitle

\section{Introduction}

Recently, Dvali and Gomez proposed a microscopic picture of geometry, and in particular of black holes, in terms of Bose-Einstein condensates (BECs) of soft gravitons \cite{Dvali:2012wq, Dvali:2} (see also \cite{Dvali:2013eja, Dvali:3, Brustein, Wintergerst-at-al-and-Berkhahn-et-al, Casadio:2013hja}). Black holes (of given charge and angular momentum) are uniquely characterised by their mass, or equivalently their Schwarzschild radius. In the present picture also the occupation number, $N$, can play this fundamental role as their sole characteristic. Hawking radiation has been explained as quantum depletion of gravitons from the condensate. Along those lines, in Ref.~\cite{Dvali:2013eja}, this approach has been generalized to anti de Sitter and de Sitter space. Also, the presence of an inflaton condensate has been studied and depletion properties, like enhancement, have been discussed.

All these condensates share the property that they are objects of {\it classical} extent, meaning that their size survives the classical limit $\hslash \rightarrow 0$. They can be related to the mechanism of classicalization \cite{Dvali:2010jz, Dvali:2011th}, and in this context are referred to as classicalons. This is the appropriate framework for the comparison of decay rates that we are interested in. It permits a discussion of a general class of bound objects given a classical Lagrangian description.

In this work we show that for an arbitrary admixture of any classicalizing substance to gravity that interacts with it, the depletion rate of the whole system is bounded from below by the black-hole depletion rate, if we disregard exact cancellations with gravity, which would require fine tuning anyway. As we will show, this remains true in the presence of additional dimensions.

Our results are implicit in the Bose-Einstein condensate picture of black holes combined with the classicalization framework, but we think it is worthwhile to spell them out in some detail.

First, we define our objects of study. By classicalization is meant the tendency of a theory to form large objects, called classicalons, of many soft quanta out of a few hard quanta. The idea is a bit similar to, but independent of, the string-theory result that UV divergences can be cut off by the formation of extended excited strings. Assuming that our theory including gravity has this classicalizing property, the important large objects can be understood in terms of classical fields and non-linearities of their equations of motion. The typical scaling behavior can then be derived from a balance of inertia and interaction, amounting to kinetic-energy and interaction-energy terms of the Lagrangian. In reality, the equations of motion govern the behavior, but it is easier to study the energy balance and the resulting scaling behavior has to be the same. In contrast to the four-dimensional gravitational case that has been studied previously \cite{Dvali:2012wq}, in higher dimensions it turns out to be essential to take the non-linearities of the field theory into account in the energy balance. This is precisely what the classicalon method does. Not recognizing this subtlety would lead to a mismatch to Hawking radiation. We regard this sensitivity to details and the gravitational non-linearities at the black hole horizon scale as new and strong support for the idea to regard black holes as graviton condensates.

Second, we apply the Bose-Einstein condensate picture; that such large objects, parametrized by a characteristic size $L$, are also characterised by the large number of $N$ of field quanta of a Bose-Einstein condensate. The number $N$ can be estimated by comparing the quantum concept of the localisation energy $\hslash c / L$ of a quantum to the total energy of the object. It is assumed that the $N$ quanta are weakly interacting, and then argued that this assumption is self-consistent. In more general situations, the energy of the condensate can be fragmented into different components, with $N_{i}$ quanta of the $i$'th field, one of them being the graviton field, which is always there.

The two ingredients above give 1) the size of a self-bound object scaling with its energy and 2) the number of quanta constituting this object, also scaling with its energy. To estimate the depletion rate of these objects we need the scaling of the effective coupling of the constituent quanta, which can also be found from the interaction terms of the Lagrangian.

Finally, we shall provide the arguments for the minimal depletion-rate property of standard black holes. We consider the regime where higher-order terms in the Lagrangian dominate those corresponding to gravitational interactions. In all such cases, barring fine tuning, which our approach cannot resolve, the scaling properties of classicalons dominated by such interaction terms always lead to increased depletion rates.


The above arguments are elaborated on in detail in Sec.~\ref{sec:Depletion}, with a particular focus in the depletion properties and half-life times for general condensates. In Sec.~\ref{sec:Summary-and-Discussion} we give a summary and a discussion. App.~\ref{app:Simplified-Self--Sourcing} deals with the simplified energy-balance argument and its shortcomings. Throughout this work, we will ignore all irrelevant prefactors and set $\hslash = c = 1$.

\section{Depletion}
\label{sec:Depletion}

As mentioned above, black holes can be viewed as classicalons. This is a manifestation of gravity's particular derivative self-interactions. It is the balance of precisely those with the kinetic term which determines the classicalization radius $r_{*}$, inside which non-linearities dominate.

To see how this comes about in more generality, let us take a scalar theory with the schematic canonically-normalized $(4 + m)$-dimensional action
\begin{align}
	\Scal
		&=
								\int\!\d^{4 + m}x\;
								\bigg[
									\frac{ 1 }{ 2 }\.\big( \partial \varphi \big)^{2}
									+
									\frac{ 1 }{ \mu^{\gamma} }\.
									\partial^{2k} \varphi^{n}
								\bigg]
								\, ,
								\label{eq:S-scalar-general}
\end{align}
with $\mu$ being some mass scale, and $\gamma = n ( m + 2 ) / 2 - ( 4 + m ) + 2k$.\footnote{The case of $k = 1$ gives, for the purpose of our discussion, the same scaling properties as General Relativity.} Generalizing the self-sourcing arguments of Ref.~\cite{Dvali:2011th} to $4 + m$ space-time dimensions, one can give an estimate of the classicalon radius,
\vs{-2mm}
\begin{align}
	r_{*}
		&\sim
								E^{\frac{ n - 2 }{ 4 k + m ( n - 2 ) + n - 6} }
								\; ,
								\label{eq:rstar}
\end{align}
where $E$ is the energy of the system. For the system to classicalize, we need to have that $r_{*}$ grows with $E$. Assuming that the $N$ quanta, of individual energy $\epsilon \sim 1 / r_{*}$, which constitute the classicalon interact weakly, we have $E \sim N \epsilon$. Together with Eq.~\eqref{eq:rstar} this yields
\begin{align}
	E
		&\sim
								N^{\frac{ 4 k + m ( n - 2 ) + n - 6 }{ 4 k - 2 ( m + 4 ) + ( m + 2 ) n }}
								\; .
								\label{eq:sqrts-in-terms-of-N}
\end{align}
For pure gravity one has $E \sim N^{\frac{ m + 1 }{ m + 2 }} \xrightarrow{\.m\.\ra\.0\.} \sqrt{N\.}$ in four dimensions. Not only $E$ but also $r_{*}$ is independent of $n$ if $k = 1$.

The above estimate of the classicalization radius takes non-linear effects into account that are not captured by a simplified energy-balance argument (\cf~App.~\ref{app:Simplified-Self--Sourcing}), where one calculates the energy-balance scale from the requirement of the formation of a bound state. More precisely, one equates the kinetic energy of one quantum to the collective potential this quantum experiences. The source of this potential is assumed to be localized. However, as we show in App.~\ref{app:Simplified-Self--Sourcing} this assumption breaks down in higher dimensions.

In order to calculate the dimensionless self-coupling of the above system, we consider an effective $( n = 4 )$-term ($k$ is still arbitrary), and study the scaling of the $2 \ra 2$ cross-section $\sigma$. As it has dimension $\mu^{- (m + 2)}$, we find
\begin{align}
	\sigma
		&\sim
								\frac{ 1 }{ \epsilon^{m + 2} }\.
								\Bigg[
									\bigg(
										\frac{ \epsilon }{ \mu }
									\bigg)^{\!\!\gamma}
								\Bigg]^{2}
		\sim
								\frac{ 1 }{ \mu^{2 \gamma} }\.\epsilon^{2 \gamma - m - 2}
		=
								\frac{ 1 }{ \mu^{4 k + 2 m} }\.\epsilon^{4 k + m - 2}
								\; .
								\label{eq:sigma2to2}
\end{align}
Let us check this for two familiar cases in $3+1$ dimensions: gravity and ordinary $\varphi^{4}$ theory. The former yields $\sigma_{\text{gravity}} \sim \epsilon^{2} / \mu^{4}$, while the latter gives $\sigma_{\varphi^{4}} \sim \epsilon^{-2}$.

Eq.~\eqref{eq:sigma2to2} defines the dimensionless self-coupling constant $\alpha$ via
\begin{align}
	\sigma
		&\equiv
								\frac{ \alpha^{2} }{ \epsilon^{m + 2} }
								\; ,
								\label{eq:2to2-crosssection}
\intertext{which yields\vs{-3mm}}
	\alpha
		&\sim
								\bigg(
									\frac{ \epsilon }{ \mu }
								\bigg)^{\! m + 2 k}
								\; .
								\label{eq:alpha}
\end{align}
Hence, with help of Eq.~\eqref{eq:sqrts-in-terms-of-N} which shows $N \sim \epsilon^{- ( m + 2 k )}$, we find
\begin{align}
	N\.\alpha\big( N \big)
		&\sim
								1
								\; .
								\label{eq:N-times-alpha}
\end{align}
This scaling property is called {\it critical} \cite{Dvali:2}. If $N \alpha\big( N \big) \gg 1$ one has an {\it over-critical} condensate.

With the above findings, we estimate the depletion rate $\Gamma$ as follows (\cf~Ref.~\cite{Dvali:2}). Let us first switch off gravity and focus on $2 \ra 2$ processes. There will be various scatterings with a spectrum of energies, but a fraction of order one of those with momentum transfer $\sim 1/r_*$ will emit a quantum out the condensate, causing depletion. This rate essentially consist of three factors: one that counts all possible pairs, \ie~$N( N - 1 ) \simeq N^{2}$, then $\alpha^{2}$, accounting for the probability of scattering any two particles, and furthermore the characteristic energy of a single quantum, $\epsilon( N ) \sim r_{*}^{-1} \sim N^{- 1 / ( 2 k + m )}$, of the depletion process. Hence,
\vs{-2mm}
\begin{align}
	\Gamma
		&\sim
								N^{2}\.\alpha( N )^{2}\.\epsilon( N )
		\sim
								\epsilon( N )
		\sim
								N^{-\frac{ 1 }{ 2 k + m }}
								\; .
								\label{eq:Gamma-phi}
\end{align}
Evidently, for $N \gg 1$, the general result \eqref{eq:Gamma-phi} is much larger than that of gravity, which reads $\Gamma_{g} \sim N^{- 1 / ( 2 + m )} \xrightarrow{\.m\.\ra\.0\.} N^{- 1 / 2}$. Above, due to criticality, the factor $N \alpha$ simply drops out. For over-critical condensates, however, this factor remains and is much larger than one, causing the condensate to be quite unstable. In fact, if $\log_{N}( N \alpha ) > 1 / ( 4 k + 2 m )$ the depletion rate actually {\it grows} with $N$.


Using Eq.~\eqref{eq:sqrts-in-terms-of-N}, we can express the mass $M \sim E$ of the condensate in terms of the number of its constituents. This allows us to translate the rate \eqref{eq:Gamma-phi} into the emission rate
\vs{-1mm}
\begin{align}
	\frac{ \d M }{ \d t }
		&=
								-\,
								\Delta M\;\Gamma
		\sim
								M^{- \frac{ 2 }{ m + 2 k -1 }}
								\; ,
								\label{eq:dM/dt-in-terms-of-N-d-dim}
\end{align}
where $\Delta M$ is the mass loss by emission of one quantum. As expected, for $k = 1$, this agrees with the standard semi-classical\./\.thermo-dynamic result
\vs{-1mm}
\begin{align}
	\frac{ \d M }{ \d t }
		&\sim
								\big[
									\text{Area}( M )
								\big]\;
								T_{\text{Hawking}}^{4 + m}( M )
		\sim
								M^{- \frac{ 2 }{ m + 1 }}
								\; ,
								\label{eq:dM/dt-Hawking}
\end{align}
where we have used the generalization of Hawking's result to general dimensions (\cf~\cite{Argyres:1998qn})\footnote{This result applies in the case in which the Schwarzschild radius is much smaller than the extent of the\\[-1mm] extra-dimension, which will always be assumed in the remainder of the work.\\[-4.8mm]}.

Now, in reality everything interacts with gravity, which will have an effect on $\Gamma$. If the system contains many more other quanta than gravitons (\ie~$N \gg N_{g} \gg 1$) these quanta will 
mainly scatter through their specific self-interaction, and deplete according to Eq.~\eqref{eq:Gamma-phi}. Thus for all condensates that are not solely constituted by gravitons, the depletion rate is enormously large, and the system spits out quanta rapidly till $N \sim N_{g}$. At this stage, the other quanta frequently interact with gravitons, with the effect that the system may have a depletion rate which might be close to that of pure gravity, but will always be {\it at least} as large.\footnote{We ignore the possibility of exact cancellation of gravity, as it would require fine tuning.}

This leads to the statement that black holes are the slowest decaying condensates with unrestricted occupation number.\footnote{This excludes \eg~solitons where the occupation number is fixed by the value of the gauge coupling, \cf~\cite{Dvali:2012wq}.} The underlying mechanism is actually the same that makes gravity the most efficient classicalizer \cite{Dvali:2011th}, and leads to the following hierarchy
\begin{align}
	\Gamma_{\text{black hole}}
		&\ll
								\Gamma_{\text{general classicalon}}
		\ll
								\Gamma_{\text{over-critical condensate}}
								\; .
								\label{eq:depletion-hierachy}
\end{align}

The rate \eqref{eq:Gamma-phi} can easily be integrated to give the half-life time $\tau$ of the $(3 + m)$-dimensional condensate. Here we define this time as the time at which half of the gravitons have been depleted away. This yields
\vs{-2mm}
\begin{align}
	\tau
		&\sim
								N^{-\frac{2 ( k - 1 )}{(2 + m) (2 k + m)}}\,
								\tau_{\text{black hole}}
								\; ,
								\label{eq:Half-life-time}
\end{align}
where we expressed the half-life time in terms of that of a corresponding black hole with the same number of constituents.

For illustrative purposes, Tab.~\ref{tab:Lifetimes} shows half-life time estimates for a condensate with the same number of quanta as a corresponding black hole of about the mass of the moon ($\approx 10^{23}\.{\rm kg}$), in units of the age of the Universe. Of course, in realistic set-ups and for large objects, the decay rates will also depend on parameters related to the geometry of the extra dimensions. As already discussed above, we see that higher values than $k = 1$ lead to a much more rapid decay.

Defining an effective spatial extra-dimension $m_{\text{eff}}$ via
\begin{align}
	\frac{ \d M }{ \d t }
		&\sim
								M^{- \frac{ 2 }{ 2 k + m - 1 }}
		\equiv
								M^{- \frac{ 2 }{ m_{\text{eff}}\.+ 1 }}
								\; ,
								\label{eq:meff-definition}
\end{align}
we find that for $k = ( m_{\text{eff}} - m + 2 ) / 2$ any integer value of $m_{\text{eff}}$ can be generated. This degeneracy can beautifully be observed in Tab.~\ref{tab:Lifetimes}. In particular for $m = 0$ a value of $k = 3$ corresponds to $m_{\text{eff}} = 4$. Furthermore, micro-condensates, of \eg~$10\.{\rm TeV}$ mass, might behave similarly to higher-dimensional black holes, with distinct collider signatures.

Note that we also included non-integer powers of $k$. Such values may arise in certain effective actions. These would be very interesting, as those objects can be reasonably long-lived, 
and may be of astrophysical importance with possibly interesting observable features. 

\begin{table}
	\begin{center}
	\begin{tabular}{|l||l|l|l|l|l|}
		\hline
		\backslashbox{\q$m$}{$k^{^{^{^{^{^{^{^{^{}}}}}}}}}$\;}
					&\qq1\;				
					&\q\;1.5\;
					&\qq2\;
					&\q\;2.5\;
					&\qq3\;\\
		\hline\hline
		\q0			&\;$10^{+33}\.\tau_{\text{H}}\vphantom{_{_{_{_{1}}}}}\vphantom{^{^{^{^{.}}}}}$
					&\;$10^{+2}\.\tau_{\text{H}}\vphantom{_{_{_{_{1}}}}}\vphantom{^{^{^{^{.}}}}}$
					&\;$10^{-8}\.\tau_{\text{H}}\vphantom{_{_{_{_{1}}}}}\vphantom{^{^{^{^{.}}}}}$
					&\;$10^{-14}\.\tau_{\text{H}}\vphantom{_{_{_{_{1}}}}}\vphantom{^{^{^{^{.}}}}}$
					&\;$10^{-17}\.\tau_{\text{H}}\vphantom{_{_{_{_{1}}}}}\vphantom{^{^{^{^{.}}}}}$\\
		\hline
		\q1			&\;$10^{+2}\.\tau_{\text{H}}\vphantom{_{_{_{_{1}}}}}\vphantom{^{^{^{1}}}}$
					&\;$10^{-8}\.\tau_{\text{H}}\vphantom{_{_{_{_{1}}}}}\vphantom{^{^{^{1}}}}$
					&\;$10^{-14}\.\tau_{\text{H}}\vphantom{_{_{_{_{1}}}}}\vphantom{^{^{^{1}}}}$
					&\;$10^{-17}\.\tau_{\text{H}}\vphantom{_{_{_{_{1}}}}}\vphantom{^{^{^{1}}}}$
					&\;$10^{-19}\.\tau_{\text{H}}\vphantom{_{_{_{_{1}}}}}\vphantom{^{^{^{1}}}}$\\
		\hline
		\q2			&\;$10^{-8}\.\tau_{\text{H}}\vphantom{_{_{_{_{1}}}}}\vphantom{^{^{^{1}}}}$
					&\;$10^{-14}\.\tau_{\text{H}}\vphantom{_{_{_{_{1}}}}}\vphantom{^{^{^{1}}}}$
					&\;$10^{-17}\.\tau_{\text{H}}\vphantom{_{_{_{_{1}}}}}\vphantom{^{^{^{1}}}}$
					&\;$10^{-19}\.\tau_{\text{H}}\vphantom{_{_{_{_{1}}}}}\vphantom{^{^{^{1}}}}$
					&\;$10^{-20}\.\tau_{\text{H}}\vphantom{_{_{_{_{1}}}}}\vphantom{^{^{^{1}}}}$\\
		\hline
		\q3			&\;$10^{-14}\tau_{\text{H}}\vphantom{_{_{_{_{1}}}}}\vphantom{^{^{^{1}}}}$
					&\;$10^{-17}\.\tau_{\text{H}}\vphantom{_{_{_{_{1}}}}}\vphantom{^{^{^{1}}}}$
					&\;$10^{-19}\.\tau_{\text{H}}\vphantom{_{_{_{_{1}}}}}\vphantom{^{^{^{1}}}}$
					&\;$10^{-20}\.\tau_{\text{H}}\vphantom{_{_{_{_{1}}}}}\vphantom{^{^{^{1}}}}$
					&\;$10^{-21}\.\tau_{\text{H}}\vphantom{_{_{_{_{1}}}}}\vphantom{^{^{^{1}}}}$\\
		\hline
		\q4			&\;$10^{-17}\tau_{\text{H}}\vphantom{_{_{_{_{1}}}}}\vphantom{^{^{^{1}}}}$
					&\;$10^{-19}\.\tau_{\text{H}}\vphantom{_{_{_{_{1}}}}}\vphantom{^{^{^{1}}}}$
					&\;$10^{-20}\.\tau_{\text{H}}\vphantom{_{_{_{_{1}}}}}\vphantom{^{^{^{1}}}}$
					&\;$10^{-21}\.\tau_{\text{H}}\vphantom{_{_{_{_{1}}}}}\vphantom{^{^{^{1}}}}$
					&\;$10^{-22}\.\tau_{\text{H}}\vphantom{_{_{_{_{1}}}}}\vphantom{^{^{^{1}}}}$\\
		\hline
	\end{tabular}
	\end{center}
	\caption{Approximate half-life times of a condensate with the same number of quanta as a corresponding black hole
			of about the mass of the moon in units of the age of the Universe for various spatial extra-dimensions
			and values of $k$. Here the $( 3 + 1 )$-dimensional Planck mass has been used.\\[2mm]}
	\label{tab:Lifetimes}
\end{table}

\section{Summary and Discussion}
\label{sec:Summary-and-Discussion}

In this work we have investigated properties of general $D$-dimensional classicalons. Generalizing the quantum $N$-portrait, originally formulated for graviton condensates in three spatial dimensions in Refs.~\cite{Dvali:2012wq, Dvali:2}, to generic classicalizing substances in arbitrary spatial dimensions, we estimated their depletion rates and derived half-life times. We found that generic classicalons are critical, and asymptotically stable for large mass. Furthermore, their depletion is enhanced as compared to the standard gravitational case. This in particular shows that black holes are the slowest decaying such objects with unrestricted occupation number, if we disregard the possibility of exact cancellation of gravity, which would require fine-tuning.

Although one motivation for the present work was simply to generalise $D = 4$ results to higher dimensional black holes, a potentially more significant result is the realisation that the simplified treatments of in Refs.~\cite{Dvali:2012wq, Dvali:2} in $D = 4$ is inadequate in the general case. As detailed in the Appendix \ref{app:Simplified-Self--Sourcing}, our full consideration of non-linearities along the lines of the classicalon discussion of Ref.~\cite{Dvali:2011th} is required to capture the right dynamical picture more generally. This application of the classicalon mechanism to decay rates is new and we thus found it to be necessary. Furthermore, the important role of the classicalon concept means that our treatment of non-gravitational classicalons can have additional applications.

A crucial property of the classicalon picture is that it fixes the scaling of the size with energy correctly in any higher dimension. Once the length scale is fixed, the wavelength of the Bose-Einstein condensate gravitons is also determined, and, in keeping with the condensate picture, we can view the black hole effectively as a coherent state of the graviton field. Indeed we have confirmed the classicalon scaling and the coherent-state picture directly in high-energy scattering \cite{Kuhnel:2014xga}. We believe that the General Relativity scaling properties of the gravitational classicalons that we have investigated ensure that their thermodynamic behaviour is that of black holes, not only with respect to the decay rates discussed in this paper, but it would be interesting if the thermodynamics could be derived purely by classicalon arguments.

Large, massive classicalons are of potential importance both in colliders and in cosmology only thanks to the property we have established: If they are produced, they evaporate more slowly with increasing size, so that sufficiently large objects can stay around long enough to lead to observable signatures.

To exemplify the above statements, we calculated the half-life times of classicalons of astrophysical mass in various dimensions and for multiple values of model parameters. We found that general condensates are generically much more short-lived than corresponding black holes, but can, for some parameter ranges, exist for cosmological time scales. Furthermore, we showed that there is a degeneracy of extra dimensions and effective higher-dimensional operators, in the sense that there are certain classicalons which have half-times corresponding to those of black holes in $\Rbbm^{(1, 3 + m)}$ for some $m$. More precisely, for any number $m$ of spatial extra dimensions, there is an effective classicalizing ($3 + 1$)-dimensional action with the same depletion properties. Although the interpretation of those generalized classicalons is not yet fully clear, we believe that they might have important physical consequences.

It is clear from our results that, everything else being equal, objects in lower dimensions are more stable asymptotically than objects in higher dimensions. In fact, what counts is evaporation out of the object, \ie~into transverse dimensions. Thus, more transverse dimensions mean more rapid evaporation. We have considered spherically-symmetric objects, but less symmetric cases may also play a role. Our discussion then indicates that \emph{evaporation} of an elongated object is suppressed relative to the evaporation of a spherical object, since the elongated object effectively has fewer transverse dimensions, at least in extreme cases. Long after the production of classicalons, we would then guess that asymmetric remnants are favoured after a substantial amount of evaporation. There are however competing mechanisms. Due to the Gregory-Laflamme instability \cite{Gregory:1993vy}, it is expected that the purely \emph{classical} evolution of black strings, \ie~extremely elongated black holes, will lead to a breakdown in the elongated direction and the production of a chain of more symmetric black holes. In contrast to evaporation, such a classical instability should persist in the limit of asymptotically large objects. So, although sustained evaporation is liable to break spherical symmetry, we are justified in paying attention to spherical objects as starting points of the evaporation process, after faster classical instabilities have produced them.

\acknowledgements
It is a pleasure to thank Gia Dvali, Daniel Flassig, Stefan Hofmann, Edvard M{\"o}rtsell, Alexander Pritzel, and Nico Wintergerst for stimulating discussions and important remarks. This work was supported by the Swedish Research Council (VR) through the Oskar Klein Centre.

\appendix
\section{Simplified Self-Sourcing}
\label{app:Simplified-Self--Sourcing}

In the true classicalon calculation of self-sourcing, the non-linearities of the field equations are important inside the classicalon radius $r_{*}$, but in fortunate cases a simplified argument is possible, and gravity in four dimensions is one example. Below we give such an argument for general power-law potentials, including gravitational potentials in higher dimensions. The argument should be handled with caution, since it assumes that the source can effectively be regarded as completely localised. This simplification actually goes wrong if the potential is too singular, leading to a strong dependence on short-distance details. Temporarily ignoring this problem we check what such a treatment would lead to.

We assume that a bound state has a dimensionless self-coupling with the energy dependence
\vs{-2mm}
\begin{align}
	\alpha
		&\sim
								\bigg(
									\frac{ \epsilon }{ \mu }
								\bigg)^{\! \kappa_{1}}
								\; ,
								\label{eq:alpha-general-over-critical-system}
\end{align}
for some $\kappa_{1} > 0$, and furthermore the effective binding potential ($\kappa_{2} > 0$)
\begin{align}
	V( r )
		&\sim
								-\;
								\mu^{1 - \kappa_{2}}\.
								\frac{ \alpha }{ r^{\kappa_{2}} }
								\; .
								\label{eq:V-Newton-d-dim}
\end{align}

Following Ref.~\cite{Dvali:2} we ask for the $N$-dependence, $N \gg 1$, of the wavelength for which the constituents form a self-sustained bound state of a certain size $L$. The corresponding equilibrium condition can be obtained from equating the kinetic energy each single quantum has, $\epsilon \sim 1 / L$, with the energy of the collective binding potential \eqref{eq:V-Newton-d-dim},
\begin{align}
	\epsilon
		&\sim
								N\.\frac{ \mu^{1 - \kappa_{1} - \kappa_{2}} \epsilon^{\kappa_{1}}}{ L^{\kappa_{2}} }
								\; .
								\label{eq:}
\end{align}
Thus, we find
\begin{align}
	L\.\mu
		&\sim
								N^{\frac{ 1 }{ \kappa_{1} + \kappa_{2} - 1 }}
								\; ,
								\label{eq:L(N)}
\end{align}
and the mass $M$ of the condensate, given approximately by the sum of the constituent's energies, becomes
\vs{-2mm}
\begin{align}
	M
		&\sim
								\mu\.N^{\frac{ \kappa_{1} + \kappa_{2} - 2 }{ \kappa_{1} + \kappa_{2} - 1 }}
								\; .
								\label{eq:M-d-dim}
\end{align}
Furthermore,
\begin{align}
	N\.\alpha( N )
		&\sim
								N^{\frac{ \kappa_{2} - 1 }{ \kappa_{1} + \kappa_{2} - 1 }}
								\; .
								\label{eq:alpha(N)}
\end{align}
The case of $\kappa_{1} = 2$ and $\kappa_{2} = 1$ reproduces that of ($3 + 1$)-dimensional gravity, and in particular $N \alpha( N ) \sim 1$. If $\kappa_{2} > 1$, we would have $N \alpha( N ) \gg 1$, \ie~an over-critical system. We would find the depletion rate
\vs{-1mm}
\begin{align}
	\Gamma
		&\sim
								N^{\frac{ 2\.\kappa_{2} - 3 }{ \kappa_{1} + \kappa_{2} - 1 }}
		\gg
								\Gamma_{\text{black hole}}
								\; ,
								\label{eq:Gamma-Simplified-Self-Sourcing}
\end{align}
in higher dimensions if the calculation was reliable. However, the assumption of a collective localized binding potential \eqref{eq:V-Newton-d-dim} is not accurate when $\kappa_{2} > 1$. In fact, for $k = 1$ and $m > 0$ we find
\vs{-2mm}
\begin{align}
	L
		&\sim
								N^{\frac{ 1 }{ 2 m + 2 }}
		\ll
								N^{\frac{ 1 }{ m + 2 }}
		\sim
								r_{*}
								\; ,
								\label{eq:L-ll-rstar}
\end{align}
which shows that the bound-state scale $L$ is always much smaller than the classicalon size. This explains the discrepancy between the depletion-rate results and the failure of the above assumptions for too singular potentials (\cf~Ref.~\cite{LL:volume-3}). It is the intrinsic non-linearity captured by the classicalon calculation that regularizes the singularity.
\newpage


\end{document}